\documentclass[sn-mathphys-ay]{sn-jnl}% Math and Physical Sciences Author Year Reference Style
\usepackage{graphicx}%
\usepackage{multirow}%
\usepackage{amsmath,amssymb,amsfonts}%
\usepackage{amsthm}%
\usepackage{mathrsfs}%
\usepackage[title]{appendix}%
\usepackage{xcolor}%
\usepackage{textcomp}%
\usepackage{manyfoot}%
\usepackage{booktabs}%
\usepackage{algorithm}%
\usepackage{algorithmicx}%
\usepackage{algpseudocode}%
\usepackage{listings}%
\usepackage{caption}
\usepackage{makecell}
\usepackage{natbib}
\theoremstyle{thmstyleone}%
\theoremstyle{thmstyletwo}%

\theoremstyle{thmstylethree}%

\begin{document}

\title[Graph Attention Network]{Graph Attention Network-Based Detection of Autism Spectrum Disorder}

%%%%%%%%%% Commented here %%%%%%%%%%%%
\author[1]{\fnm{Abigail} \sur{Kelly}\textsuperscript{[0009-0009-6850-468X]}}\email{amk7r@mtmail.mtsu.edu}

\author*[2]{\fnm{Ramchandra} \sur{Rimal}\textsuperscript{[0000-0001-8182-0593]}}\email{ramchandra.rimal@mtsu.edu}

\author[3]{\fnm{Arpan} \sur{Sainju}\textsuperscript{[0000-0001-5668-194X]}}\email{arpan.sainju@mtsu.edu}

% % \author[1,2]{\fnm{Third} \sur{Author}}\email{iiiauthor@gmail.com}
% % \equalcont{These authors contributed equally to this work.}

\affil[1]{\orgdiv{Computational and Data Science}, \orgname{Middle Tennessee State University}, \orgaddress{\street{1301 E. Main St.}, \city{Murfreesboro}, \postcode{37132}, \state{TN}, \country{USA}}}

\affil*[2]{\orgdiv{Department of Mathematical Sciences}, \orgname{Middle Tennessee State University}, \orgaddress{\street{1301 E. Main St.}, \city{Murfreesboro}, \postcode{37132}, \state{TN}, \country{USA}}}

\affil[2]{\orgdiv{Department of Computer Science}, \orgname{Middle Tennessee State University}, \orgaddress{\street{1301 E. Main St.}, \city{Murfreesboro}, \postcode{37132}, \state{TN}, \country{USA}}}

%%%%%%%%%% End Commented part here %%%%%%%%%%%%
% \affil[3]{\orgdiv{Department}, \orgname{Organization}, \orgaddress{\street{Street}, \city{City}, \postcode{610101}, \state{State}, \country{Country}}}


\abstract{Autism Spectrum Disorder (ASD) is a neurodevelopmental condition characterized by atypical brain connectivity. One of the crucial steps in addressing ASD is its early detection. This study introduces a novel computational framework that employs an Attention-Based Graph Convolutional Network, referred to as the GATGraphClassifier, for detecting ASD. We utilize Functional Magnetic Resonance Imaging (fMRI) data from the Autism Brain Imaging Data Exchange (ABIDE) repository to construct functional connectivity matrices using Pearson correlation, which captures interactions between various brain regions. These matrices are then transformed into graph representations, where the nodes and edges represent the brain regions and functional connections, respectively. The GATGraphClassifier employs attention mechanisms to identify critical connectivity patterns, thereby enhancing the model's interpretability and diagnostic accuracy. Our proposed framework demonstrates superior performance across all standard classification metrics compared to existing state-of-the-art methods. Notably, we achieved an average accuracy of 88.79\% on the test data over 30 independent runs, surpassing the benchmark model's performance by 12.27\%. In addition, we identified the crucial brain regions associated with ASD consistent with the previous studies and a few novel regions. This study not only contributes to the advancement of ASD detection but also shows the potential for broader adaptability of GATGraphClassifier in analyzing complex relational data in various fields, where understanding intricate connectivity and interaction patterns is essential.}

\keywords{brain imaging, ASD, GAT, fMRI, GCN}

\maketitle

\section{Introduction}
Autism Spectrum Disorder (ASD) is a neurodevelopmental condition characterized by challenges in social communication, repetitive behaviors, and restricted interests. Its prevalence has steadily increased, making early detection and intervention critically important~\citep{lord2018autism}. According to estimates from the CDC's Autism and Developmental Disabilities Monitoring Network, about 1 in 36 children have been identified with ASD. Research indicates that early diagnosis can significantly improve developmental outcomes, enabling targeted therapies during crucial stages of brain development~\citep{kohli2022role, bivarchi2021barriers}. Emerging evidence points to disruptions in brain connectivity patterns as a hallmark of ASD, making these patterns a promising avenue for diagnostic tools~\citep{belmonte2004autism, o2017functional}. One of the emerging tools for early detection of ASD is Functional Magnetic Resonance Imaging (fMRI).

fMRI is a widely used non-invasive neuroimaging technique to capture brain activity over time. It operates by detecting fluctuations in the blood-oxygen-level-dependent (BOLD) signal, which reflects changes in neural activity within small volumetric units called voxels at distinct time points. Consequently, fMRI data is structured as a time series, representing the dynamic activity of each voxel throughout the scanning ~\citep{canario2021review}. Among its various applications, resting-state fMRI (rs-fMRI) is particularly valuable in studying brain disorders, as it enables the examination of intrinsic functional connectivity patterns while the subject is not engaged in any task. This approach has been instrumental in early detection and identifying biomarkers for neurological and psychiatric conditions, including ASD, schizophrenia, and Alzheimer's disease~\citep{feng2022review, almuqhim2021asd, chelladurai2023fmri, wang2023deep, algumaei2022feature}.

Recent advancements in graph-based neural networks, particularly Attention-Based Graph Convolutional Networks (GCNs), offer a novel approach to detecting ASD. The use of attention mechanisms allows these models to focus on the most relevant connections within the brain's functional network, potentially improving diagnostic accuracy~\citep{gu2023autism, yin2021graph}. This approach is particularly suited for capturing the complex, non-linear relationships inherent in functional connectivity data, which traditional methods may overlook.

In this study, we leverage fMRI data to analyze the functional connectivity patterns of the brain, representing these patterns as graphs, where nodes correspond to specific brain regions and edges denote their functional relationships~\citep{stanley2013defining}. By applying Attention-Based GCNs to this graph representation, we aim to uncover subtle connectivity disruptions indicative of ASD.

Our study contributes to the field in several ways: (1) we propose a novel application of attention-based GCNs for ASD detection, (2) we evaluate the performance of this method on fMRI-based functional connectivity data significantly outperforming the previous results, (3) we explore the interpretability of attention mechanisms in identifying the most diagnostically relevant brain connections, and (4) we employ SHapley Additive exPlanations (SHAP) to investigate the associations between specific brain regions and ASD. This work aims to advance the understanding of ASD-related brain connectivity and demonstrate the potential of graph-based deep learning methods in neurodevelopmental disorder diagnosis.

\section{Related Work}

The application of machine learning in diagnosing mental disorders has gained popularity and demonstrated significant success. Over time, more advanced techniques have been developed to better capture the complex structures and relationships inherent in mental health data. The common early approaches included logistic regression (LR), support vector machine (SVM), and random forest (RF). More recently, deep learning-based methods have become more prevalent, particularly deep neural networks (DNNs), convolutional neural networks (CNNs), and recurrent neural networks (RNNs). Additionally, graph-based methods, such as Graph Neural Networks (GNNs), have gained prominence for their ability to model relationships within graph-structured data. In particular, Graph Attention Networks (GATs) have emerged as a powerful extension, utilizing attention mechanisms to assign varying levels of importance to connections within the graph. This capability allows GATs to capture complex dependencies more effectively, making them a promising approach for diagnosing mental disorders.

\subsection{Traditional Machine Learning Approaches}

The early approaches to ASD classification relied mainly on LR, SVM, RF, and various advanced techniques derived from these foundational methods~\citep{wang2024ifc}. These methods often required handcrafted feature extraction from imaging modalities such as structural MRI (sMRI), fMRI, and diffusion tensor imaging (DTI)~\citep{eslami2019asd}. For example, LR and SVM were frequently combined with feature extraction techniques to classify neuroimaging data~\citep{zafar2017classification, xie2009brain}. Some studies leveraged SVM to analyze spatial and temporal aspects of fMRI data~\citep{song2009unsupervised}, achieving improved results at the single-subject level~\citep{mourao2006impact}. However, while existing methods rely on temporal and spatial feature extraction techniques, they often fail to fully integrate these features into the model development process~\citep{wang2024ifc}. This limitation hindered their ability to comprehensively capture the complex patterns inherent in brain connectivity data.

\subsection{Deep Learning}

The advent of deep learning has facilitated the development of advanced methodologies capable of automatically extracting hierarchical features from neuroimaging data. CNNs and RNNs are increasingly being employed for disease classification using neuroimaging data. In particular, studies utilizing CNNs for classifying fMRI data have gained significant attention due to their ability to autonomously extract meaningful features~\citep{horikawa2017generic, meszlenyi2017resting}. RNNs are particularly valued for their efficacy in capturing long-term dependencies within sequential data~\citep{khan2024aff, rimal2024identifying, rimal2023comparative}. These methods have been applied not only for the identification of ASD, but also to differentiate various stages of Alzheimer’s disease~\citep{jo2019deep, zhao2023conventional, arya2023systematic} and to detect schizophrenia~\citep{zhang2023detecting, zheng2021diagnosis}. 

In addition, hybrid approaches that combine these models with other techniques have gained prominence as they have the potential to enhance performance. For example, CNNs are being used in conjunction with SVMs~\citep{nie20163d}, as well as with autoencoders~\citep{huang2017modeling}, auto-decoder networks~\citep{wen2018neural}, and 3D convolutional autoencoders~\citep{zhao2017constructing}. RNNs are currently being explored for the early detection of Alzheimer’s disease and are also being utilized in conjunction with autoencoders for the classification of Attention-Deficit/Hyperactivity Disorder (ADHD) from fMRI data~\citep{qiang2021modeling, ismail2024use}. Furthermore, CNNs are under active investigation for their potential to detect a range of brain disorders~\citep{lin2022convolutional}. While these methods demonstrate the potential to learn hierarchical feature representations directly from fMRI data, they struggle to incorporate the inherent graph structure of brain connectivity data, leading to the exploration of graph-based approaches. We discuss the work on graph-based methods in more detail in the following two sections.

\subsubsection{Graph Neural Networks}
GNNs provide a natural and powerful framework for modeling brain connectivity data, which is typically represented as a graph \( G = (V, E) \), where \( V \) denotes the set of brain regions (nodes), and \( E \) represents the functional or structural connections (edges) between them. Connectivity strength can be encoded as edge weights, forming an adjacency matrix \( A \), while node attributes can capture regional activity features derived from fMRI or other neuroimaging modalities.  

GNNs have shown promise in various neuroimaging applications. For instance, they are being used to predict brain age in patients with Alzheimer's disease based on rsfMRI data~\citep{gao2023brain, chen2024comparative}. Furthermore, recent advancements have adapted GNNs for fMRI analysis, incorporating task-aware brain connectivity~\citep{yu2022learning}, modeling spatiotemporal dynamics in resting-state fMRI data~\citep{azevedo2022deep}, and investigating topological properties of GNNs in fMRI-based applications~\citep{pitsik2023topology, mohammadi2024graph}.
Additionally, graph representation learning techniques are proving effective in identifying ASD patients through graph embedding methods~\citep{yousefian2023detection}. Other research focuses on leveraging GNNs for predictive modeling by capturing the 3D spatial structures present in rsfMRI data~\citep{jiang2022cnng}.

Despite their flexibility in capturing complex graph-structured data, GNNs face limitations. One major challenge is their difficulty in accounting for the varying importance of nodes and edges within heterogeneous brain networks, which can lead to suboptimal feature extraction. To address this, recent studies have explored attention-based GNN variants that assign adaptive weights to different connections, enhancing the interpretability and predictive power of brain connectivity models.

\subsubsection{Graph Attention Networks}

GATs address the limitations of GCNs by incorporating attention mechanisms that dynamically assign learnable weights to nodes and edges, enabling more nuanced feature extraction from graph-structured data. Recent advances demonstrate GATs' effectiveness in neurodevelopmental and neurological disorder diagnosis, including ASD detection using spatial-constrained sparse functional brain networks~\citep{yang2021autism}, multimodal brain connectomics-based prediction of Parkinson's disease~\citep{safai2022multimodal}, and schizophrenia diagnosis via multi-GATs~\citep{yu2023multi}. Further innovations include relational GATs for ASD classification~\citep{gu2023autism}, spatiotemporal attention models for dynamic brain connectome representation~\citep{kim2021learning}, and adversarial learning-based node-edge GATs~\citep{chen2022adversarial}. These approaches highlight GATs' ability to model complex functional and structural relationships in brain networks, offering superior interpretability and predictive performance over traditional GCNs, by selectively attending to the most informative connections.

The benchmark approach used in this paper, ASD-SWNet~\citep{zhang2024asd}, introduces a shared-weight framework combining an autoencoder for unsupervised feature learning and a custom CNN for supervised classification. To mitigate challenges posed by small sample sizes, the model employs data augmentation and is evaluated on a preprocessed ABIDE-I dataset (N=871 subjects) using nested ten-fold cross-validation. ASD-SWNet achieves state-of-the-art results with 76.52\% accuracy and an AUC of 0.81, outperforming existing baselines. This success underscores the potential of integrating GAT-inspired attention mechanisms with hybrid deep learning architectures for improved ASD diagnosis.

\begin{figure}[htb]
    \centering
    \includegraphics[width=\linewidth]{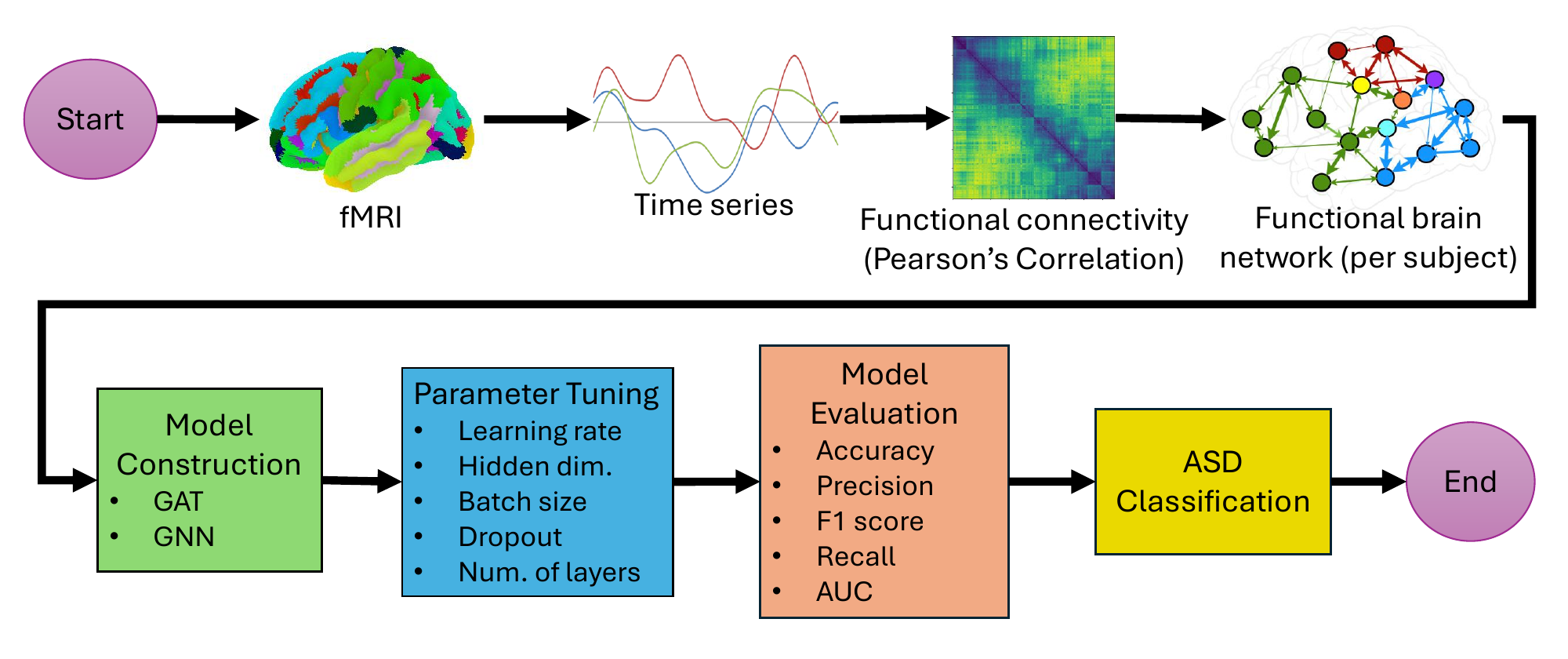}
    \caption{Flow of Overall Approach}
    \label{fig:overall-approach}
\end{figure}

\section{Methods}
The overall flow of our approach is shown in Fig.~\ref{fig:overall-approach}.  We began by collecting the fMRI data and extracting the time series.  Next, the functional connectivity was calculated using Pearson's correlation, followed by the functional brain network where a functional brain network is created per subject.  Finally, the models are constructed, tuned, and tested.  More details are explained in this section.

\subsection{Data Acquisition}
For our experiment, we use the Autism Brain Imaging Data Exchange (ABIDE) data set accessed using the Nilearn library~\citep{ABIDEdata, Nilearn}.  The ABIDE-I data set is publicly available and contains brain imaging data of 1,035 resting-state fMRI data with 505 subjects with ASD and 530 subjects that are healthy.  
The data is pooled from 17 different research institutions around the world.  We adopt the Configurable Pipeline for the Analysis of Connectomes (C-PAC) for the collection of data~\citep{craddock2013towards}.  This pipeline builds upon existing software packages, such as Analysis of Functional NeuroImages (AFNI)~\citep{cox1996afni}, FMRIB Software Library (FSL)~\citep{smith2004advances}, and Advanced Normalization Tools (ANTs)~\citep{avants2009advanced}.  This pipeline ensures standardized preprocessing, including steps like motion correction and spatial normalization.  Additionally, band-pass filtering was applied to the data to remove high-frequency noise and low-frequency drift.

For region-of-interest (ROI) analysis, we chose the Automated Anatomical Labeling (AAL) atlas~\citep{tzourio2002automated}.  The atlas divides the brain into 116 anatomically distinct regions, facilitating the extraction of representative time series from predefined areas.  To extract ROI-specific time series data, we used the NiftiLabelsMasker class from Nilearn~\citep{weber2022confounds}.  This masker maps the volumetric fMRI data onto the AAL atlas by associating each voxel with an anatomical label.  The time series is standardized by applying Z-score normalization.  The missing values in the time series are replaced with zero using SimpleImputer.

\subsection{Graph Construction}
Constructing a graph enables the integration of imaging data, such as fMRI, with non-imaging information~\citep{gu2023autism}. To achieve this, we compute functional connectivity to represent the time series data as a graph. Functional connectivity is calculated using the Pearson correlation coefficient, which is the most common method in the literature~\citep{sedgwick2012pearson, hyde2019applications, nawghare2024early}. This correlation measures the temporal coherence between pairs of regions of interest (ROIs)~\citep{linke2020dynamic}, offering valuable insights into inter-regional functional relationships within the brain.

Let $G = \{V, E, A\}$ represent an undirected connected graph, where $V$ is the set of nodes ($|V| = n$), $E$ is the set of edges, and $A$ is the adjacency matrix. The entry $A(i, j)$ in the adjacency matrix denotes whether an edge exists between nodes $i$ and $j$. Specifically, $A(i, j) = 1$ if an edge exists; otherwise, $A(i, j) = 0$~\citep{zhang2019graph}. Additionally, let a vector $x \in \mathbb R^n$ represent a graph signal defined on the nodes of $G$, where $x(i)$ corresponds to the signal value at node $i$. This signal can also be interpreted as a node attribute. We further denote $X \in \mathbb R^{n \times d}$ as the node attribute matrix of the graph, where each column in $X$ represents one of the $d$ signals associated with the graph~\citep{zhang2019graph}.
In this framework, brain regions based on the AAL atlas serve as nodes in the graph. An edge connects two nodes if a meaningful relationship exists between them. To determine connectivity, we apply a threshold of absolute correlation at 0.2. We construct the graph once the edge set for all 116 brain regions is established based on this threshold. The weights of the edges correspond to the pairwise Pearson correlation values between ROIs.

\subsection{Graph Convolution}
\begin{figure}[htb]
\centering
\begin{minipage}{.5\textwidth}
  \centering
  \includegraphics[width=.5\linewidth]{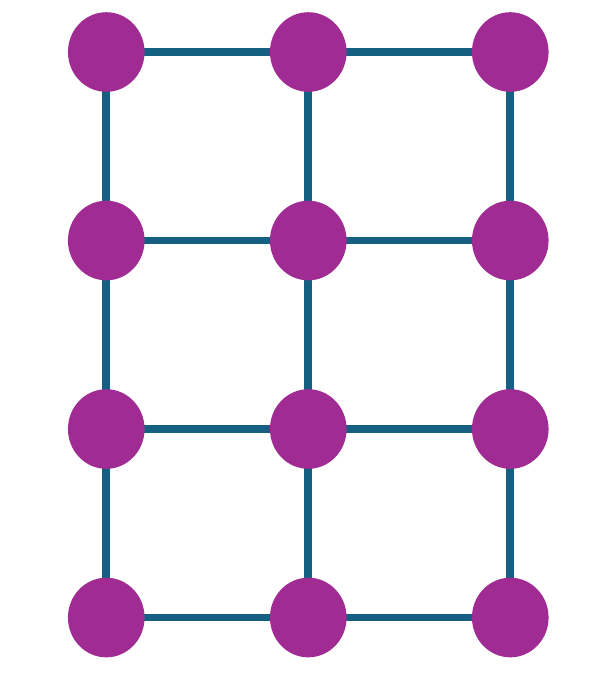}
  \caption{CNN Structure}
  \label{fig:CNN}
\end{minipage}%
\begin{minipage}{.5\textwidth}
  \centering
  \includegraphics[width=.5\linewidth]{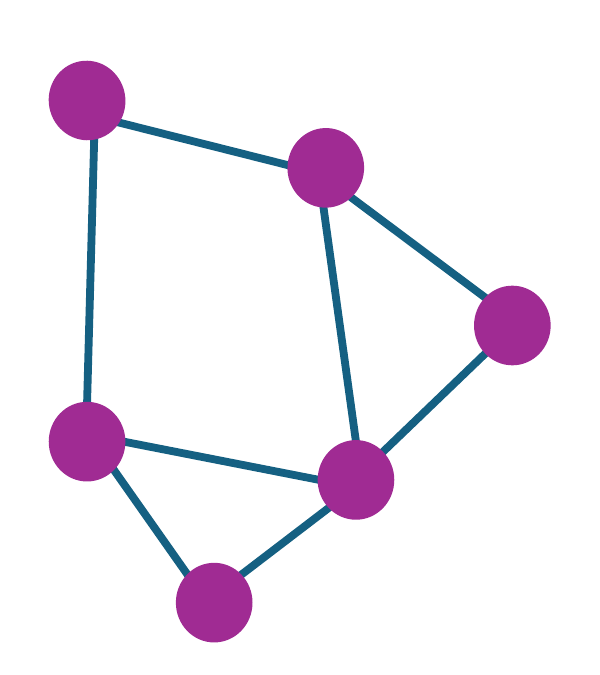}
  \caption{GCN Structure}
  \label{fig:GCN}
\end{minipage}
\end{figure}

GCNs are neural networks designed to process graph-structured data~\citep{kipf2016semi, wu2020comprehensive, zhou2020graph}. In a typical CNN, as shown in Fig.~\ref{fig:CNN}, a grid is placed over an image, and a filter is moved across it. CNNs are effective for grid-like data structures such as images but are not well-suited for data without a predefined structure.
GCNs, on the other hand, as shown in Fig.~\ref{fig:GCN}, extend the convolution operation used in CNNs to non-grid-like structures, making them appropriate for analyzing graph-based data~\citep{khemani2024review}.

In GCNs, the convolution operation involves aggregating features from a node's neighbors. During this process, each node updates its features by combining its own features with those of its neighbors, typically using a weighted sum or mean to capture the local graph structure and relationships. Following this aggregation, a transformation is applied, usually consisting of a linear layer followed by a non-linear activation function. The mathematical formulation for a single GCN layer is:
\begin{equation*}
H^{(l + 1)} = \sigma (\tilde{D}^{-1/2} \tilde{A} \tilde{D}^{-1/2} H^{(l)} W^{(l)})
\end{equation*}
where:
\begin{itemize}
\item $H^{(l)}$: Node features at layer $l$
\item $\tilde{A}$: Adjacency matrix of the graph with added self-loops ($\tilde{A} = A + I$), where $I$ is the identity matrix
\item $\tilde{D}$: Diagonal degree matrix of $\tilde{A}$
\item $W^{(l)}$: Trainable weight matrix for layer $l$
\item $\sigma$: Activation function
\end{itemize}
This formulation highlights the iterative process of aggregation and transformation of features. The inclusion of $\tilde{D}^{-1/2}$ provides normalization for the adjacency matrix $\tilde{A}$, ensuring that the contributions from a node's neighbors are appropriately scaled. This normalization helps prevent numerical instabilities during training. By stacking multiple GCN layers, the model can capture higher-order and more complex relationships by aggregating information from increasingly distant neighbors in the graph~\citep{khemani2024review}.

This design enables GCNs to learn rich, context-aware node embeddings that effectively incorporate both graph structure and node features. This capability has driven their widespread adoption across diverse applications, from social network analysis to molecular graph prediction and brain connectivity studies~\citep{khemani2024review, pitsik2023topology, mohammadi2024graph}. In neuroimaging applications, nodes typically represent ROIs while edges denote functional or structural connections, creating a natural framework for analyzing graph-structured brain data.

Despite their strengths in capturing local graph structure, GCNs face certain limitations. A significant challenge is the oversmoothing phenomenon~\citep{chen2020measuring}, where node embeddings become increasingly similar and eventually indistinguishable after stacking multiple layers. This limitation has motivated the development of several GCN variants, including GATs~\citep{velickovic2017graph} and spectral graph convolutions~\citep{wang2018local}, which aim to improve performance and address these inherent constraints.

\subsection{Attention-based GCN Mechanism}
While traditional GCNs rely on a fixed aggregation scheme (typically a weighted sum or mean) to update node features, GATs introduce an attention mechanism that allows nodes to assign different levels of importance to their neighbors. This attention mechanism adapts the aggregation process based on the relevance of each neighboring node, making the model more flexible and expressive in capturing complex relationships within the graph structure.

In GATs, instead of using a uniform aggregation function, each edge is assigned an attention coefficient, which is computed using a self-attention mechanism~\citep{velickovic2017graph}. Specifically, the attention coefficients between node $i$ and node $j$ are determined by a learned attention function $\alpha_{ij}$, which is applied to the feature vectors of the nodes. The attention coefficient is typically computed as:

\begin{equation*}
    \alpha = \frac{\text{exp(LeakyReLU(}\textbf{a}^T [W\textbf{h}_i || W\textbf{h}_j]))}{\sum_{k \in N(i)} \text{exp(LeakyReLU(}\textbf{a}^T[W\textbf{h}_i || W\textbf{h}_k]))}
\end{equation*}
where 
\begin{itemize}
    \item $W$: Weight matrix applied to node features
    \item $\textbf{h}_i$ and $\textbf{h}_j$: Feature vectors for nodes $i$ and $j$
    \item $\textbf{a}$: Learnable attention vector
    \item $||$: Concatenation operation
\end{itemize}

The attention coefficient $\alpha_{ij}$ quantifies the importance of the features of node $j$ when updating the features of node $i$. The denominator ensures that the attention scores are normalized across all neighbors of node $i$, making them comparable~\citep{vrahatis2024graph}.

Once the attention coefficients are computed, the feature updates are performed as follows:

\begin{equation*}
    H_i^{(l+1)} = \sigma\left(\sum_{j \in N(i)} \alpha_{ij} W^{(l)} \textbf{h}_j\right)
\end{equation*}

where $N(i)$ denotes the set of neighbors of node $i$, and the update equation aggregates the weighted features of neighboring nodes using the attention coefficients. This mechanism allows the model to learn a dynamic weighting scheme, assigning more importance to some neighbors while down-weighting less relevant ones~\citep{vrahatis2024graph}.

The attention mechanism offers several advantages in graph-based learning. By enabling the model to focus on the most relevant parts of the graph~\citep{vrahatis2024graph}, it effectively mitigates the impact of noisy or irrelevant neighbors. Moreover, the dynamic computation of attention coefficients allows GATs to adapt to diverse graph structures and tasks, demonstrating their versatility across a wide range of applications, including node classification, graph classification, and link prediction. The incorporation of multi-head attention further enhances the model's expressiveness by computing multiple sets of attention coefficients in parallel~\citep{cheng2021drug}. This approach enables GATs to capture a broader spectrum of relationships, thereby improving their ability to learn more informative and robust node representations.

In the context of brain connectivity analysis, GATs are particularly advantageous when it is necessary to assign varying levels of importance to different brain regions or connections. This capability allows the model to focus on the most relevant regions for tasks such as disease prediction or brain state classification. By learning adaptive weights for each connection, GATs provide a more flexible and robust solution compared to traditional GCNs, making them well-suited for analyzing complex neuroimaging data.

\subsection{Model Architecture}

In this study, we conducted an in-depth examination of two models: GCN and GAT. We systematically experimented with a range of hyperparameters, including learning rate ($1\times10^{-3}$ to $1\times10^{-6}$), hidden dimension (128, 256, 512, 1024), batch size (8, 16, 32, 64), dropout rate (0.1-0.6), the number of heads in the GAT (2, 4, 6, 8) and the number of layers (4-7 GAT/GCN layers and 1-4 dense layers) for each model.
%\textcolor{red}{It is helpful to discuss what was the range of the parameter values experimented during the tuning step.} 
 Our experimental results demonstrated that GAT outperformed GCN, thus supporting the conclusion of the literature review. As a result, all experiments presented in this paper were carried out using the GAT-based model, which we refer to as GATGraphClassifier.

The GATGraphClassifier is a GAT-based architecture specifically designed for graph-based classification tasks. It was built utilizing the networkx library and the PyTorch Geometric library built upon  PyTorch.   It consists of seven GATConv layers, each utilizing eight attention heads to effectively aggregate information from neighboring nodes. To enhance model stability and mitigate overfitting, batch normalization and dropout layers are incorporated following each GATConv layer. The final node embeddings are aggregated through global mean pooling and subsequently processed by two fully connected (FC) layers, which project the learned representations into the output space. Different activation functions are employed at various stages of the network to introduce non-linearity and facilitate effective learning. The Exponential Linear Unit (ELU) activation function is applied after each GAT layer to stabilize gradient flow, while the Rectified Linear Unit (ReLU) activation function is used in the fully connected layers. Finally, the log-softmax activation function is applied in the output layer to compute class probabilities for binary classification. The complete final model architecture is detailed in Table \ref{tab:gat_architecture}.

\begin{table}[h]
    \centering
    \caption{GAT-Based Model Architecture}
    \setlength{\tabcolsep}{10pt}
    \renewcommand{\arraystretch}{1.5} % Increased row height for spacing
    \begin{tabular}{|c|c|c|c|}
        \hline
        \textbf{Layer(s)} & \textbf{Details} & \textbf{Output Shape} & \textbf{Activation} \\
        \hline
        Input & - & (N, 316) & - \\
        GAT Block (1) & \makecell[c]{GATConv (8 heads), BatchNorm,\\ Dropout (p=0.1)} & (N, 2048) & ELU \\
        GAT Blocks (2–7) & \makecell[c]{GATConv (8 heads), BatchNorm,\\ Dropout (p=0.2)} & (N, 2048) & ELU \\
        Pooling & Global Mean Pool & (N, 2048) & - \\
        FC1 & \makecell[c]{Linear (2048 → 1024),\\ Dropout (p=0.2)} & (N, 1024) & ReLU \\
        FC2 & Linear (1024 → 2) & (N, 2) & LogSoftmax \\
        \hline
    \end{tabular}
    \label{tab:gat_architecture}
\end{table}

\begin{figure}[htb]
    \centering
    \begin{minipage}{0.5\textwidth}
        \centering
        \includegraphics[width=\linewidth]{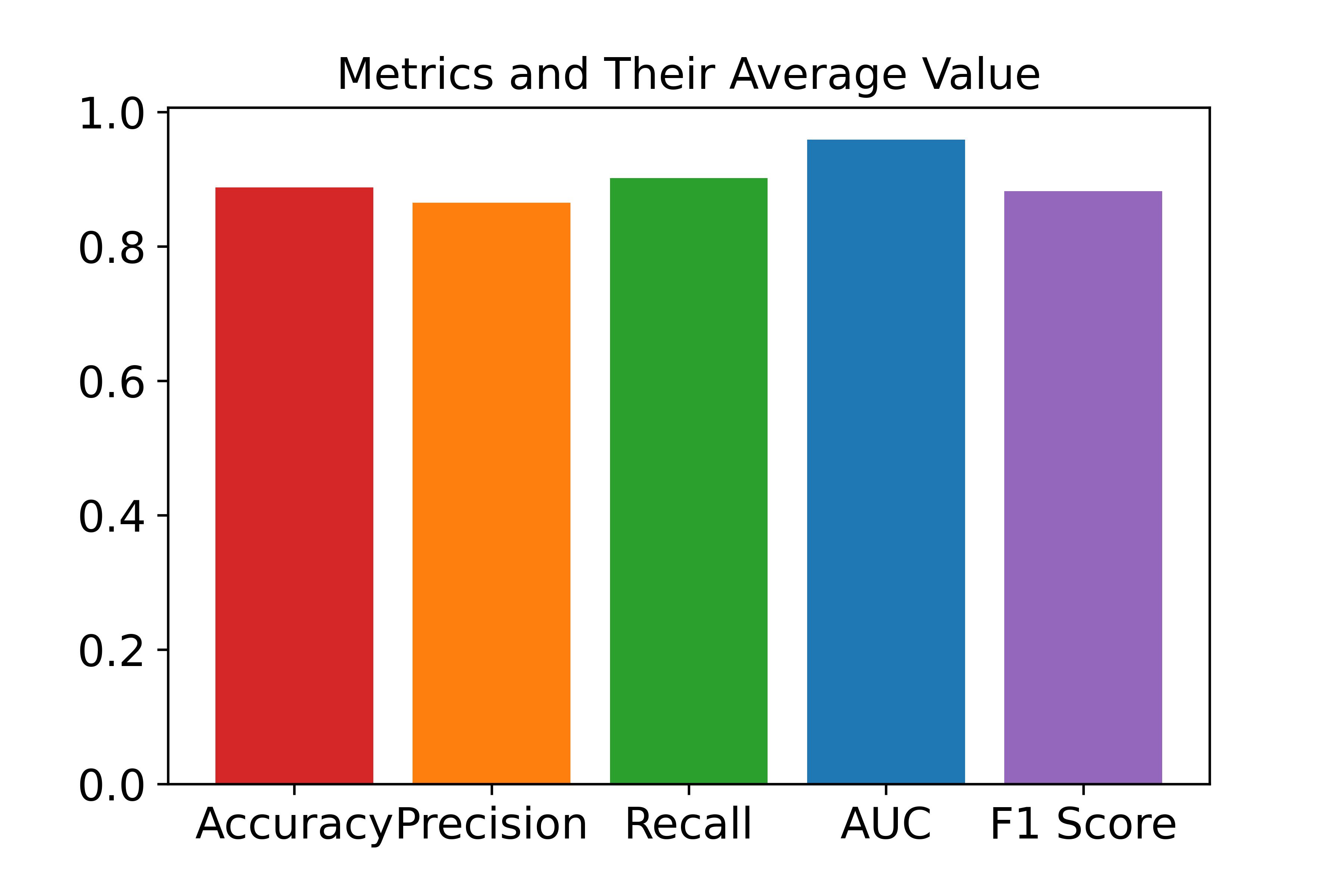}
        \caption{Average Metric Values}
        \label{fig:all_metrics_bar}
    \end{minipage}%
    \begin{minipage}{0.5\textwidth}
        \centering
        \includegraphics[width=\linewidth]{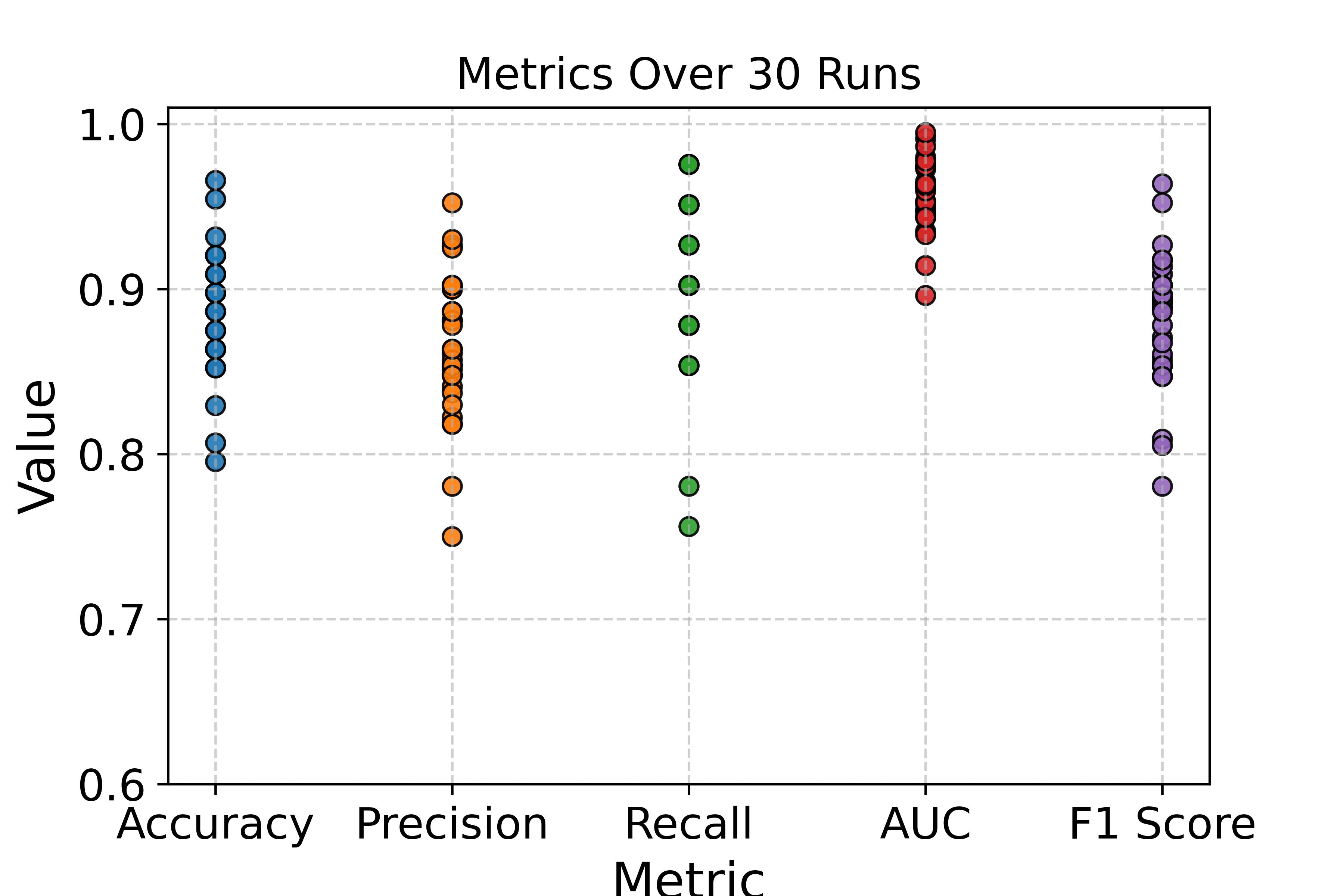}
        \caption{Metric Values over Runs}
        \label{fig:all_metrics_line}
    \end{minipage}
\end{figure}

\begin{figure}[htb]
    \centering
    \begin{minipage}{0.5\textwidth}
        \centering
        \includegraphics[width=\linewidth]{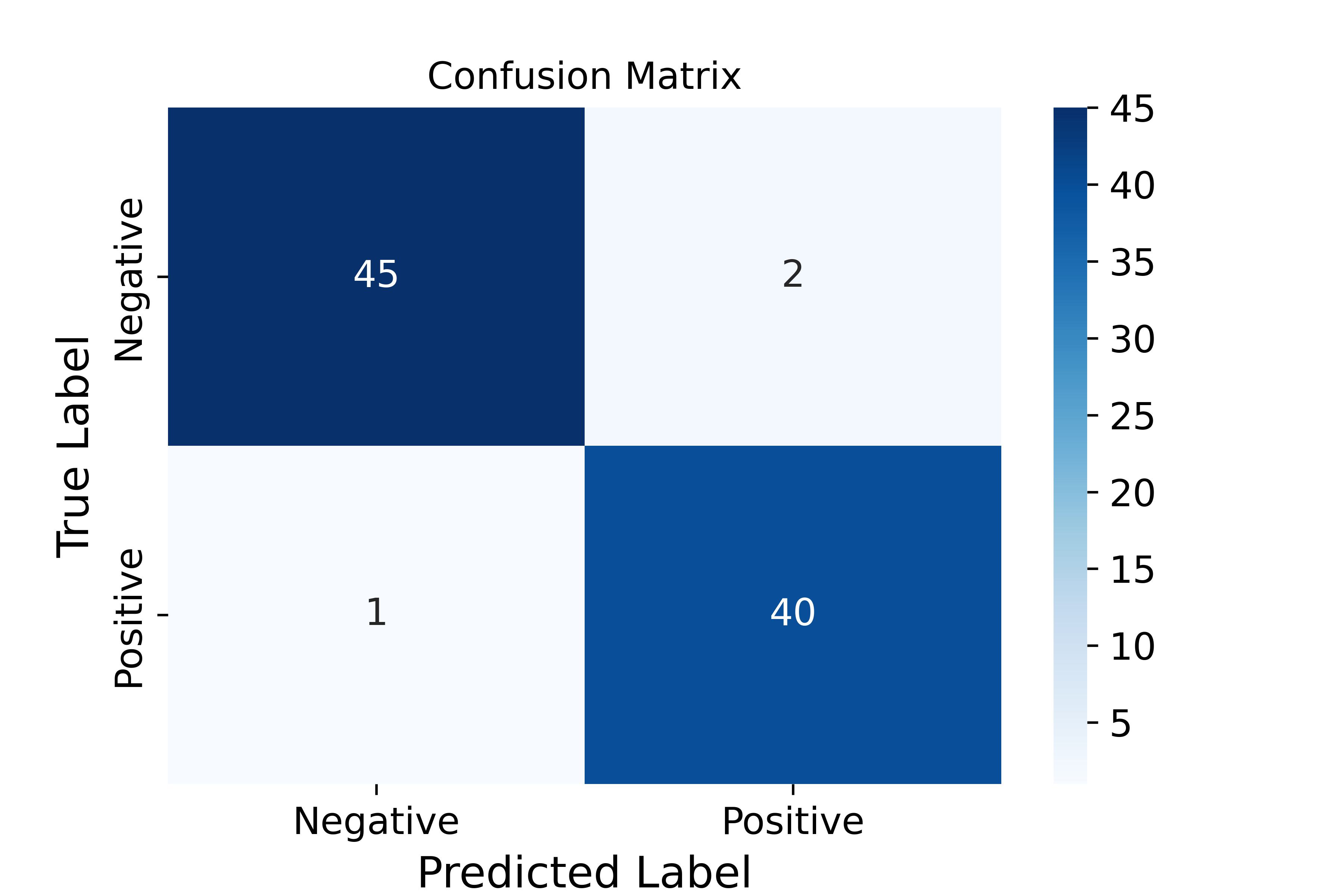}
        \caption{Confusion Matrix of Best Run}
        \label{fig:confusion_matrix}
    \end{minipage}%
    \begin{minipage}{0.5\textwidth}
        \centering
        \includegraphics[width=\linewidth]{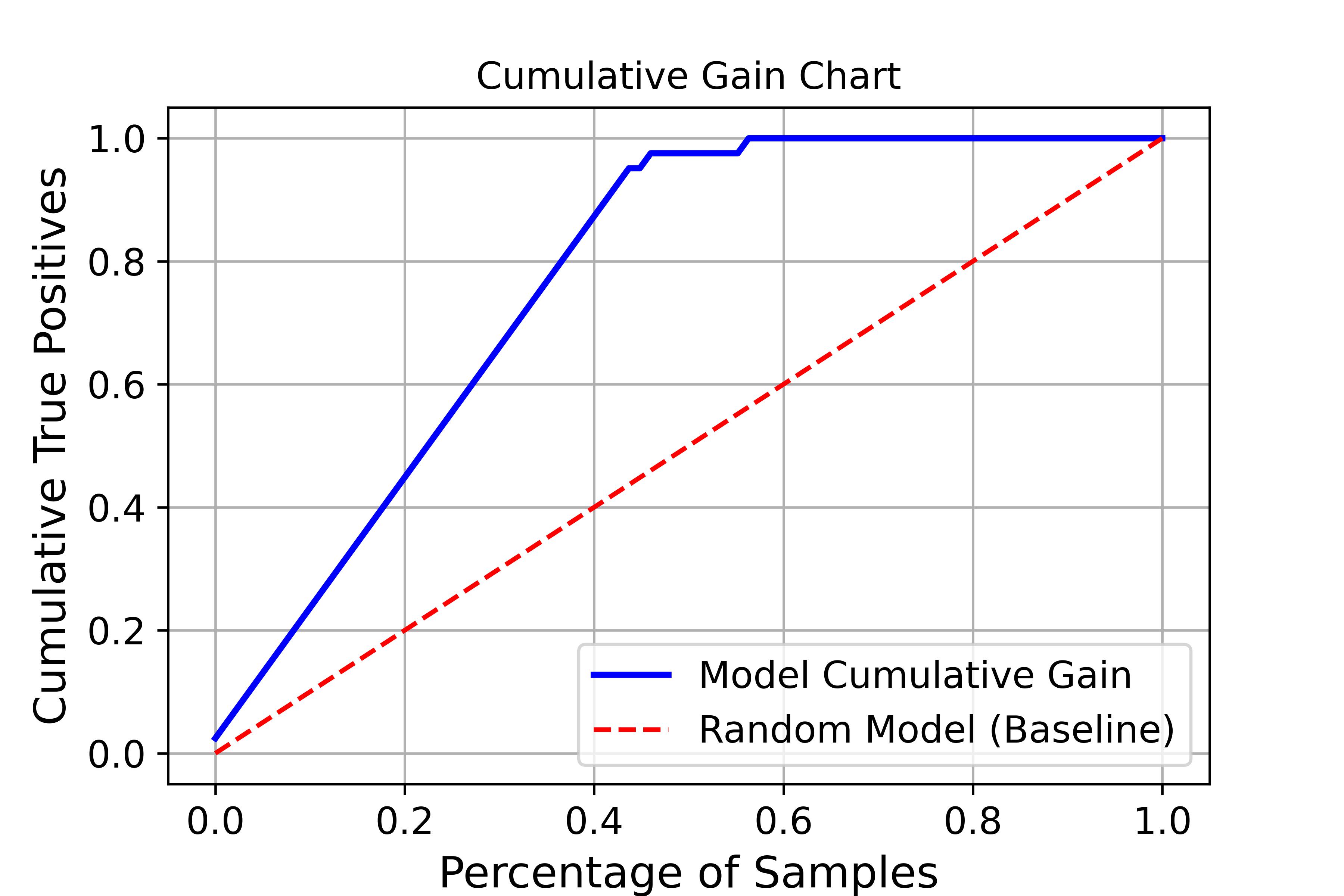}
        \caption{Cumulative Gain of Best Run}
        \label{fig:cumulative_gain}
    \end{minipage}
\end{figure}

\begin{figure}[htb]
    \centering
    \begin{minipage}{0.5\textwidth}
        \centering
        \includegraphics[width=\linewidth]{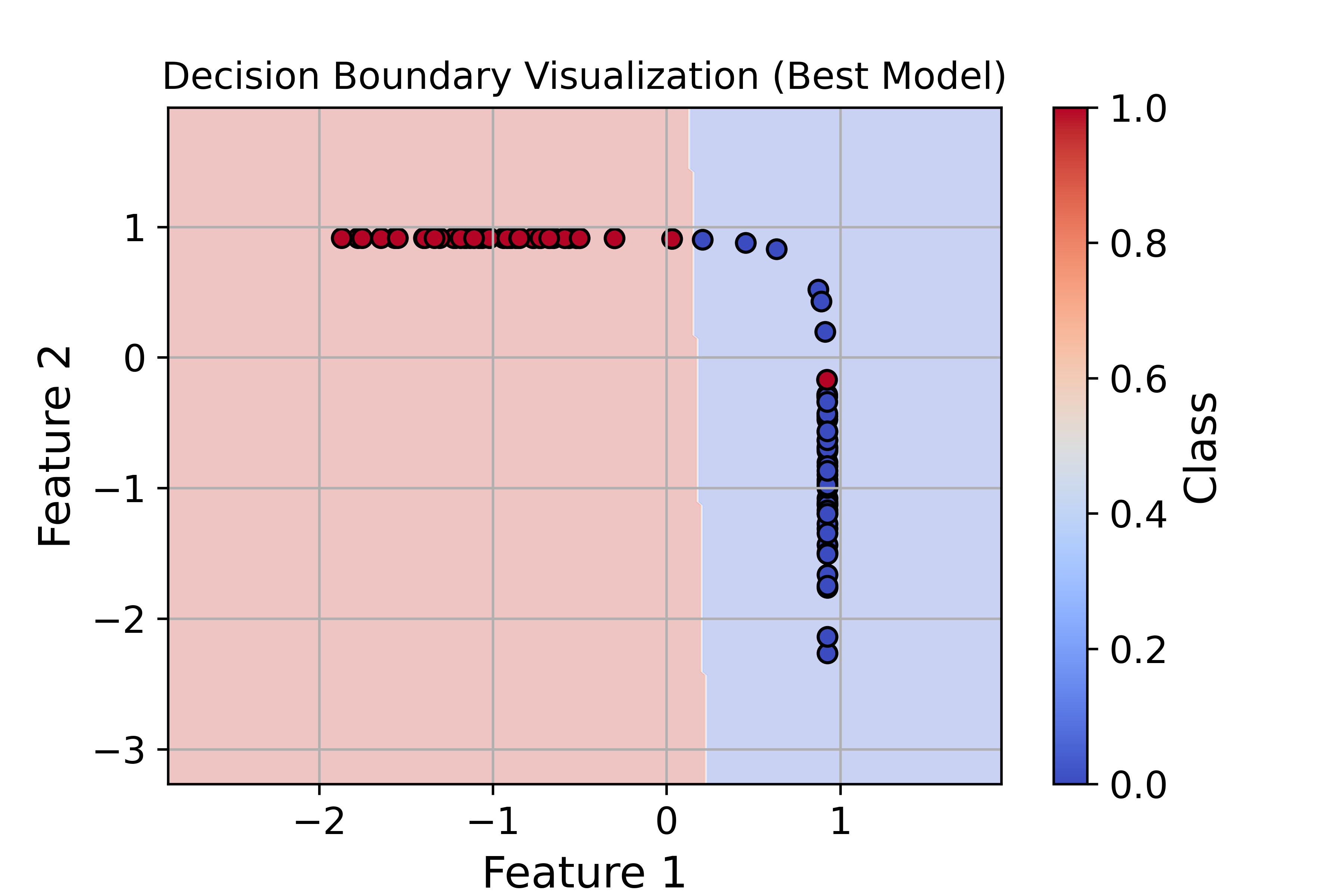}
        \caption{Decision Boundary for Best Run}
        \label{fig:decision_boundary}
    \end{minipage}%
    \begin{minipage}{0.5\textwidth}
        \centering
        \includegraphics[width=\linewidth]{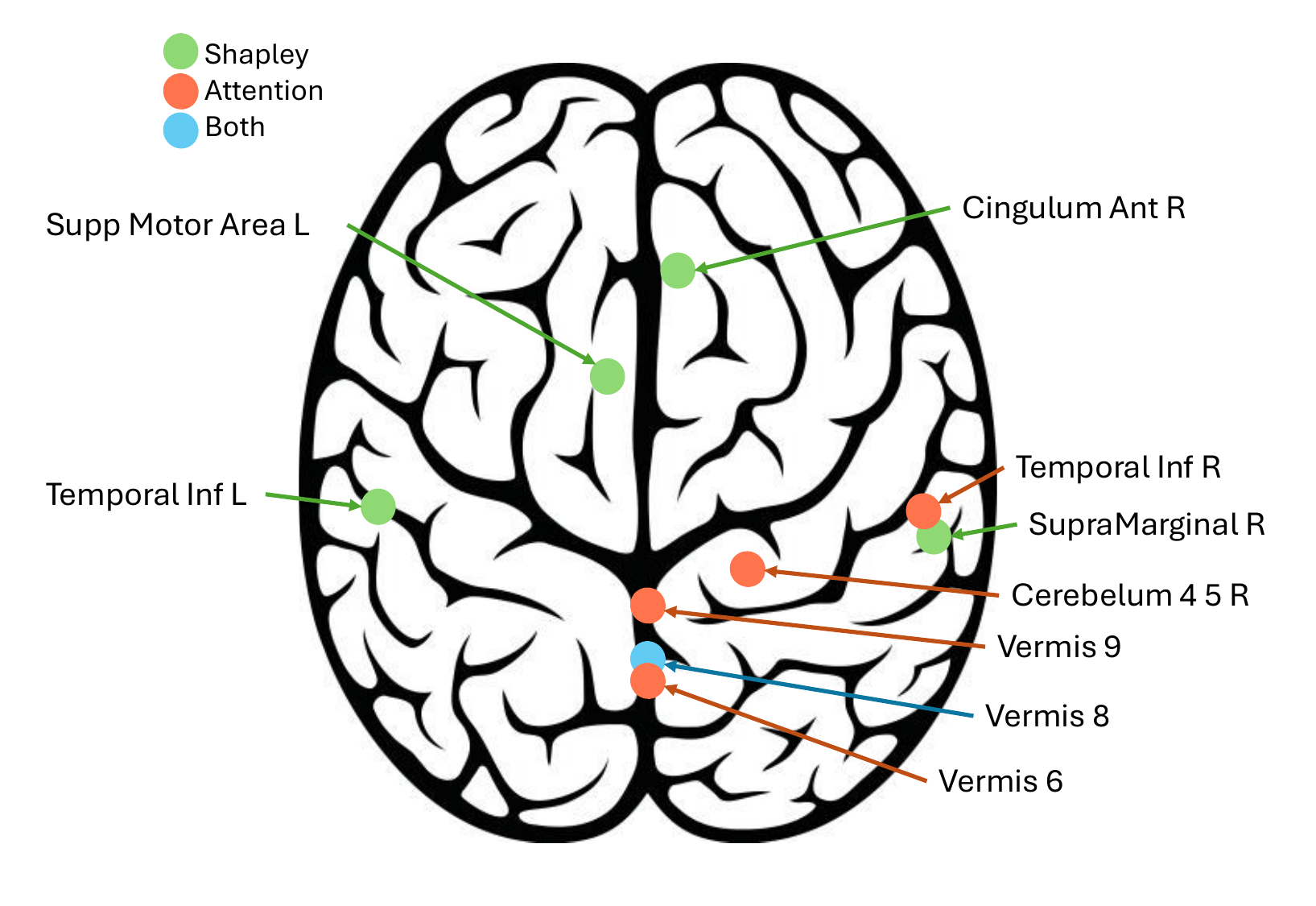}
        \caption{Most Useful Brain Regions}
        \label{fig:brain_regions}
    \end{minipage}
\end{figure}

\section{Results}
In this study, we examined the effectiveness of our proposed method by conducting a comparative evaluation against the benchmark state-of-the-art model using the ABIDE-I dataset.  While we did not make any attempts to replicate the results published in the comparative articles, our findings represent the average performance scores achieved on test data,  a dataset that was not utilized during the training phase of our model.   Specifically, we allocated 80\% of the dataset for training purposes, 10\% for validation, and the remaining 10\% for testing.     

The performance results presented in this work are based on the average metrics obtained over 30 independent testing runs. In each run, the dataset was randomly split into training, validation, and testing sets using stratified sampling to preserve class distribution. The model was reloaded and evaluated on a new test set each time, and the standard classification metrics such as accuracy, precision, recall, F1 score, and area under the curve (AUC) were recorded. We report the mean and standard deviation of these metrics across all 30 runs to ensure a robust and reliable evaluation. These scores are displayed in Table~\ref{tab:performance_over_runs} and visually represented in Fig.~\ref{fig:all_metrics_bar}. We analyzed the performance across all the metrics listed above, reporting their average, minimum, maximum, and standard deviations across the replicates. For a more detailed understanding of performance across different iterations, the metrics from each individual run are illustrated in Fig.~\ref{fig:all_metrics_line}. We observe that there is a slightly high variation in the precision and recall scores, 
 followed by the F1 score and accuracy, and the lowest variation in the AUC score. However, the lowest performance score across the replicates was still higher than the best performance we observed in the literature review. These results underscore the superior classification capabilities of our model.

The maximum accuracy achieved by our model was 96.59\%. The confusion matrix corresponding to this run is presented in Fig.~\ref{fig:confusion_matrix}. In this instance, the model produced one false positive, indicating that one normal subject was incorrectly classified as an ASD subject, and two false negatives, meaning two ASD subjects were mistakenly classified as normal, among a total of 88 subjects in the test data. The model correctly classifies the remaining 85 subjects. Additionally, the cumulative gain for the best-performing run of our model is illustrated in Fig.~\ref{fig:cumulative_gain}. The plotted line represents the cumulative gain in correctly detecting fMRI images associated with ASD. Notably, the model achieves maximal cumulative true positives at a significantly faster rate compared to the random baseline model. Furthermore, it is observed that after processing approximately 58\% of the samples, the accuracy plateaus, indicating no further improvement.

The decision boundary of the best-performing model was also analyzed. Fig.~\ref{fig:decision_boundary} illustrates the decision boundary for this run, generated using t-SNE applied to the embeddings extracted from the model's layer preceding the classification layer. This visualization provides insight into how the model organizes graph embeddings in the feature space. The results demonstrate that the model effectively separates the two classes, with only minor misclassifications observed.

The performance of our model, compared to the benchmark approach and other existing representative methods, is summarized in Table~\ref{tab:performance_based_on_existing_models}. Additionally, the average, minimum, maximum, and standard deviation of performance metrics across 30 replicates are presented in Table~\ref{tab:performance_over_runs}. The results indicate that the accuracy scores of our model range from 0.79 to 0.97, precision ranges from 0.75 to 0.95, recall from 0.75 to 0.98, AUC from 0.90 to 0.99, and the F1 score ranges between 0.78 and 0.96. Looking at average performance scores, the recall score of 90.16\% shows that, on average, our model successfully identifies 90.16\% cases of ASD among all patients with ASD in the test dataset. Furthermore, when our model predicts that a patient has ASD, it is accurate approximately 87\% of the time on average. Finally, across all subjects, our model achieves an average accuracy of approximately 89\% in classifying individuals as either ASD or non-ASD.

While there is modest variation in performance across replicates, even the scores within two standard deviations below the mean outperform the benchmark used in this study. Specifically, our best accuracy is 95.59\%, the average accuracy is 88.79\%, and the worst-case accuracy is 79.55\%. Notably, even the worst-case accuracy of our model exceeds the average accuracies reported for other models in Table~\ref{tab:performance_based_on_existing_models}. Additionally, our worst recall and AUC values surpass those of all other approaches, while the worst precision of our model outperforms all models except for ASD-SWNet, which achieves a slightly higher precision. When comparing the average performance metrics of our model to those reported for other approaches, our approach significantly outperforms all comparable methods across all metrics. The performance metrics for comparison with other models listed in Table~\ref{tab:performance_based_on_existing_models} are sourced from~\citep{zhang2024asd}.

Finally, Table~\ref{tab:scores} presents the top five most important brain regions identified through SHAP scores and attention mechanism scores. SHAP assigns each feature an importance value for a particular prediction~\citep{lundberg2017unified}. The SHAP scores were computed by first flattening the feature matrix of each subject into a one-dimensional vector. A wrapper prediction function was then employed to reshape the vector back into the original feature matrix size, after which the features were aggregated using the mean before being fed into the GAT. The KernelExplainer from the SHAP library was used to reduce the background dataset and compute SHAP values for the flattened input. For each brain region, the average SHAP value was calculated to quantify its contribution to the model's predictions.
Our analysis revealed that only 9 out of 116 brain regions exhibited nonzero SHAP values, indicating that these nine regions are the primary contributors to our model's decision-making process. Among these, the top five brain regions ranked by SHAP value are Vermis 8, Cingulum Ant R, Temporal Inf L, SupraMarginal R, and Supp Motor Area L.

The attention mechanism scores were obtained from the attention weights calculated during the evaluation of the test set for each GAT convolution layer. Each GAT layer was configured with \texttt{return\_attention\_weights=True}, which allowed for the extraction of attention weights during the model evaluation process. For each of the 30 runs, the attention weights across multiple attention heads were averaged to generate a per-edge coefficient. Node importance was subsequently determined by summing the incoming attention values for all edges connected to a specific node. Based on this analysis, the top five brain regions ranked by attention weights were Vermis 6, Vermis 8, Vermis 9, Temporal Inf R, and Cerebellum 4 5 R.

The top brain regions identified using the SHAP score and attention weights are different, as shown in Fig.~\ref{fig:brain_regions}.  This difference is not unexpected due to the fact that the two methods capture different aspects of our model's behavior.  The SHAP values measure the impact of perturbing each region's features on the final output, while the attention weights measure where the model is looking or what the model is focusing on at specific layers.  These different perspectives can yield different results, so caution must be used when interpreting results, as multiple interpretability methods should be used to gain a fuller understanding of model decisions.

Many of the important brain regions identified by the proposed model agree with alterations reported in previous studies.  These include functional abnormalities in cerebellar vermis lobules~\citep{courchesne2013hypoplasia, hashimoto1995development, webb2009cerebellar, riva2013gray, khan2015cerebro}, reduced connectivity involving the right anterior cingulate cortex (Cingulum Ant R)~\citep{assaf2010abnormal, cherkassky2006functional}, altered connectivity involving the inferior temporal regions associated with facial recognition and language processing (Temporal Inf L and R)~\citep{cheng2015autism, cai2018increased, lim2015disorder, foster2015structural}, atypical connectivity patterns within the supramarginal gyrus (SupraMarginal R) related to social cognition~\citep{gotts2012fractionation, maximo2013approaches}, disruptions in supplementary motor area (Supp Motor Area L) connectivity relevant to motor planning and repetitive behaviors~\citep{cheng2015autism, mellema2022reproducible}, and altered sensorimotor connectivity involving cerebellar lobules 4 and 5 (Cerebelum 4 5 R)~\citep{cakar2024functional, mellema2022reproducible}. Notably, our model revealed novel brain regions associated with ASD that have not been highlighted in prior literature. These regions warrant further investigation and may contribute to a deeper understanding of the pathophysiology of ASD.

\begin{table}[]
    \centering
    \setlength{\tabcolsep}{10pt}
    \renewcommand{\arraystretch}{1.2}
    \caption{Performance Metrics of our Model}
    \begin{tabular}{|l|l|l|l|l|}
    \hline
        \textbf{Metric} & \textbf{Average} & \textbf{Minimum} & \textbf{Maximum} & \textbf{Standard Deviation} \\
        \hline
        Accuracy & 88.79 & 79.55 & 96.59 & 3.72 \\
        Precision & 86.50 & 75.00 & 95.24 & 4.34 \\
        Recall & 90.16 & 75.61 & 97.56 & 5.02 \\
        AUC & 0.96 & 0.90 & 0.99 & 0.02 \\
        F1 Score & 0.88 & 0.78 & 0.96 & 0.04 \\
        \hline
    \end{tabular}
    \label{tab:performance_over_runs}
\end{table}

\begin{table}[]
    \centering
    \setlength{\tabcolsep}{10pt}
    \renewcommand{\arraystretch}{1.2}
    \caption{Performance of our Model vs. Existing Models (from~\citep{zhang2024asd})}
    \begin{tabular}{|l|l|l|l|l|}
        \hline
        \textbf{Method} & \textbf{Accuracy} & \textbf{Precision} & \textbf{Recall} & \textbf{AUC} \\
        \hline
        ASD-DiagNet~\citep{eslami2019asd} & 69.07 & 65.25 & 69.58 & 0.71 \\
        ASD-SWNet~\citep{zhang2024asd} & 76.52 & 76.15 & 80.65 & 0.81 \\
        Hi-GCN~\citep{jiang2020hi} & 72.63 & 65.16 & 70.54 & 0.79 \\
        GCN~\citep{parisot2018disease} & 70.58 & 68.32 & 74.29 & 0.72 \\
        Our Approach & \textbf{88.79} & \textbf{86.50} & \textbf{90.16} &  \textbf{0.96}\\
        \hline
    \end{tabular}
    \label{tab:performance_based_on_existing_models}
\end{table}

\begin{table}[]
    \centering
    \caption{Top 5 Most Important Brain Regions based on Shapley Score and Attention Mechanism Score}
    \begin{tabular}{|l|l|}
    \hline
         \textbf{Shapley (with Score)}& \textbf{Attention (with Score)}  \\
         \hline
         Vermis 8 ($3.294 \times 10^{-4}$) & Vermis 6 (3.706) \\
         Cingulum Ant R ($3.292 \times 10^{-4}$) & Vermis 8 (3.658) \\
         Temporal Inf L ($2.004 \times 10^{-4}$) & Vermis 9 (3.625) \\
         SupraMarginal R ($1.947 \times 10^{-4}$) & Temporal Inf R (3.558) \\
         Supp Motor Area L ($1.800 \times 10^{-4}$) & Cerebelum 4 5 R (3.444)\\
    \hline
    \end{tabular}
    \label{tab:scores}
\end{table}

\section{Discussion}
The findings of this study highlight the effectiveness of graph-based models—particularly attention-based GCNs—for the early detection of ASD using fMRI data. We found that integrating attention mechanisms significantly improves the model's capability to prioritize crucial regions and connections within the brain graph, allowing for a more effective focus on the most pertinent features for classification. Our proposed framework, GATGraphClassifier, outperforms current state-of-the-art methods across all key performance metrics, demonstrating its superiority and potential for clinical application.

A key advantage of attention-based GCNs is their interpretability, which is a key consideration in clinical applications. This is achieved by assigning attention weights to specific regions and connections between regions in the brain. This capability enables the model to capture subtle yet meaningful patterns in brain connectivity associated with ASD, thereby improving both classification accuracy and explanatory power. Notably, GATGraphClassifier also identified novel brain regions, specifically Frontal Sup R and Frontal Inf Tri R, that have not been commonly reported in prior ASD research. While these findings require further validation, they represent promising avenues for future investigation and may contribute to biomarker discovery and a deeper understanding of ASD neurobiology.

Despite these strengths, a notable limitation of GATGraphClassifier lies in the increased computational complexity introduced by attention-based GCNs, which may hinder scalability for large datasets or real-time clinical applications. Future work will focus on optimizing the model to enhance efficiency and facilitate broader deployment in clinical settings.

\section{Conclusion}

In this study, we proposed a novel application of attention-based GCNs for detecting ASD using fMRI data. The attention mechanism enables nodes to assign varying levels of importance to their neighbors, enhancing the model's flexibility and expressiveness in capturing complex relationships within the brain connectivity graph.

We thoroughly experimented with GCN and GAT models by systematically tuning key hyperparameters, including learning rate, hidden dimensions, batch size, dropout rate, number of attention heads, and model depth. Our results demonstrate that GATGraphClassifier achieves state-of-the-art performance in ASD classification across all performance metrics, with an average accuracy of 88.79\%, outperforming the benchmark model by 12.27\%. 
Beyond classification performance, we explored the statistical interpretability of the models by analyzing attention scores and SHAP values. This analysis identified several brain regions consistent with previous studies and novel regions that may warrant further investigation, offering new directions for ASD research and potential biomarker discovery.
Moreover, our computational framework is both customizable and user-friendly, making it suitable for analyzing fMRI data related to other neurological disorders such as Alzheimer's disease and schizophrenia. The broader utility of attention-based GCNs extends to domains including healthcare, neuroscience, social network analysis, urban planning, and finance. GCNs hold significant promise in advancing ASD diagnosis and understanding by improving diagnostic accuracy, facilitating biomarker identification, offering insights into ASD pathophysiology, and supporting the goals of precision medicine.

\section*{Declarations}

\begin{itemize}
    \item \textbf{Funding:} {\small Ramchandra Rimal and Arpan Sainju were partially supported by MTSU Data Science Institute Seed Grant.}
    \item \textbf{Conflict of interest/Competing interests:} The authors have no conflict of interest to declare that are relevant to the content of this article.
    \item \textbf{Ethics approval and consent to participate:} Not Applicable.
    \item \textbf{Consent for publication:} Not Applicable.
    \item \textbf{Data availability:} {\small All data used in this study are derived from publicly available datasets.}
    \item \textbf{Materials availability:} Not Applicable.
    \item \textbf{Code availability:} {\small Codes will be made available on GitHub to the public once the manuscript is accepted for publication.}
    \item \textbf{Author contribution:} {\small R.R., A.K., and A.S.  conceptualized the research problem and methodology. R.R collected and prepared the data. A.K.  worked on implementing models and wrote the initial draft. R.R. revised the draft, and A.S. reviewed the final draft. All authors contributed to the figures,  the final review, and the editing of the manuscript.}
\end{itemize}

\bibliography{sn-bibliography}% common bib file

\appendix
\section{Appendix}
Results of the GCN that we tested.  These results are the average metrics from 30 runs using the same data set and split as our GAT approach.

\begin{table}[h]
    \centering
    \setlength{\tabcolsep}{10pt}
    \renewcommand{\arraystretch}{1.2}
    \caption{Performance metrics of GCN model}
    \begin{tabular}{|l|l|l|l|l|}
    \hline
        \textbf{Metric} & \textbf{Average} & \textbf{Minimum} & \textbf{Maximum} & \textbf{Standard Deviation} \\
        \hline
        Accuracy & 46.45 & 42.05 & 51.17 & 2.40 \\
        Precision & 46.27 & 43.90 & 48.78 & 1.30 \\
        Recall & 94.23 & 87.80 & 100.00 & 2.92 \\
        AUC & 0.46 & 0.29 & 0.62 & 0.07 \\
        F1 Score & 0.62 & 0.59 & 0.96 & 0.02 \\
        \hline
    \end{tabular}
    \label{tab:gcn_over_runs}
\end{table}

\end{document}